# Simulations on Consumer Tests: A Perspective for Driver Assistance Systems

Delf Block
Volkswagen AG
Development Car Safety
Electronic and Testing
Wolfsburg, Germany

Sönke Heeren
Automotive Safety
Technologies GmbH
Kompetenzcluster Wolfsburg
Ingolstadt, Germany

Stefan Kühnel
Volkswagen AG
Development Car Safety
Electronic and Testing
Wolfsburg, Germany

André Leschke
Volkswagen AG
Development Car Safety
Electronic and Testing
Wolfsburg, Germany

Bernhard Rumpe
RWTH Aachen University
Software Engineering
Aachen, Germany

Vladislavs Serebro
Automotive Safety
Technologies GmbH
Kompetenzcluster Wolfsburg
Ingolstadt, Germany

## ABSTRACT
This article discusses new challenges for series development regarding the vehicle safety that arise from the recently published AEB test protocol by the consumer-test-organisation EuroNCAP for driver assistance systems [6]. The tests from the test protocol are of great significance for an OEM that sells millions of cars each year, due to the fact that a positive rating of the vehicle-under-test (VUT) in safety relevant aspects is important for the reputation of a car manufacturer. The further intensification and aggravation of the test requirements for those systems is one of the challenges, that has to be mastered in order to continuously make significant contributions to safety for high-volume cars. Therefore, it is to be shown how a simulation approach may support the development process, especially with tolerance analysis. This article discusses the current stage of work, steps that are planned for the future and results that can be expected at the end of such an analysis.

## 1. INTRODUCTION AND MOTIVATION

The variety of advanced driver assistance systems is steadily increasing in the area of comfort and is making significant contributions within the area of safety. This progress has been taken into account by consumer-test-organizations like EuroNCAP, by planning to rate the performance of driver assistance systems such as seatbelt reminder or electronic stability control as opposed to giving a fixed rating on whether it is a standard option in the vehicle or not. From the start of 2014, newly released vehicles are going to be tested and assessed under quality aspects similiar to those tests of passive safety features. That way, more driver assistance systems like forward collision warning (FCW) and automated emergency braking (AEB) will be examined for their abilities of preventing and mitigating crashes.

On the website of EuroNCAP the test procedures are explicitly described in an officially published test protocol, where conditions and parameters of the tests are only being tolerated within specific ranges. It is being assumed that a test procedure will be carried out only once. However, the complexity of scenarios in the field and the state-of-the-art technology of identifying surrounding objects, do require an investigation of possible tolerance ranges of the system's performance during these test scenarios and how such ranges will influence an assessment later on.

In reality, these tolerance investigations can only be reproduced with utmost effort. Autonomous steering and braking robots as well as a very precise differential GPS can help to a certain degree of accuracy within a cm resolution. The installation of the robots into the VUTs and the preparation of the testing ground will take a lot of time and effort. Closed-loop simulations however, may allow an intensive exploration of influencing parameters and also may ensure test reproducibility. Moreover, several parameters may be investigated by their influence on the overall results simultaneously and under controllable conditions.

This paper discusses, which opportunities may arise by looking into the basis of signal processing simulations and which benefit these may have for series development of driver assistance systems, hereby taking into account the equivalence class partitioning (ECP) method used in general software testing. This principle consciously reduces the number of tests for a particular range of input data by choosing a single, representative input.

The remainder of the article is organized as follows: Section 2 gives a brief overview of related works that have influenced this paper. Section 3.1 explains the testing procedures of AEB City and AEB Inter-Urban published by EuroNCAP. Then, in section 3.2 a method is outlined, how the development of a simulation environment may be systemized. Section 3.2.1 to 3.2.3 reveals the concrete implementation of this methodology. In section 3.3, some aspects of using the simulation environment for the analysis of tolerance ranges are discussed and what kind of knowledge should be obtained by it in the future.

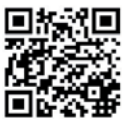

## 2. RELATED WORK

In the last few years, a number of papers wrt. simulation environments and other virtual testing procedures were published. In the following, only a selection of publications that influenced the different development steps of this environment are mentioned. Berger et al. already developed a methodology of virtual acceptance testing which was successfully applied in the DARPA Urban Challenge. Especially the aspects of concentrating on analyzing the circumstances of the system under development (SUD), the design and use of a Domain Specific Language (DSL) for the system's context and the derivation of metrics had an influence on this work [1, 3]. In order to test different types of ADAS, Dirndorfer, v. Neumann-Cosel et al. also designed a simulation framework, whose main simulation components and experiences, i.e. "Virtual Test Drive" (VTD)[19] and "Automotive Data- and Time-triggered Framework" (ADTF) were transferred to the here outlined approach [4, 12, 13]. The framework differs from Berger's aforementioned who designed an environment entirely from scratch. While Berger et al. and Dirndorfer et al. concentrated on a software-in-the-loop (SIL) setup, v. Neumann-Cosel et al. outlined this toolset with the ability of addressing various in-the-loop methods, i.e. driver- or vehicle-in-the-loop [11]. Nentwig et al. enhanced the simulation environment for other hardware-in-the-loop setups to test several vehicle functions combined with the real hardware. Parts of their results were applied to the series development of Volkswagen Enterprise [10]. The here outlined approach rely on a software solution in order to be able to abstract from real hardware and realtime requirements.

In addition, some other papers are referred that also focus on simulations of ADAS using different types of software tools. Martinus et al. made use of a similar simulation methodology by deploying other components and tools to realize a hardware-independent solution, that is able to integrate software without specific target hardware [9]. Another HIL-simulation approach is outlined by Schlager et al., trying to reduce the complexity of realizing such environments. At the same time, this approach ensures the scalability of the interfaces for the realization of an integrated architecture, concentrating on HIL-specific aspects [17]. Tideman et al. present a simulation environment by connecting PreSCAN and dSPACE components with each other to test a lane keeping assist [18]. All approaches have in common, that they enhanced HIL environments for a more practical and generic use. The here presented approach tries to abstract from real hardware.

## 3. SIMULATIONS ON CONSUMER TESTS

Hereafter, the focus is on the upcoming challenges in the new testing procedures for series development, thereby initially describing prime aspects of EuroNCAP's test scenarios. Following that, a method is outlined which may help designing simulation environments more systematically by concentrating on their intended purpose of later usage. Here, the environment's purpose is considered in supporting the definition and fulfillment of requirements. The tolerance analysis of driver assistance systems performing in the particular test cases with possibly varying parameters and boundary conditions play a significant role.

### 3.1 EuroNCAP's AEB Test Protocol

In the recent version of EuroNCAP's AEB Test Protocol from July 2013, the process of test runs "AEB City" and "AEB Inter-Urban" was presented. From the start of 2014, it will be necessary for the vehicle to be equipped with an active safety system preventing or mitigating collisions with low or intermediate speed to achieve a 5-star rating. The assessment particularly concentrates on quality aspects, i.e. how well a driver assistance system performs in the given test scenarios and not only evaluates their abscence or presence in the vehicle.

There are three different types of test scenarios (see figure 1):

- Car-To-Car-Rear: stationary (CCRs)
- Car-To-Car-Rear: mobile (CCRm)
- Car-To-Car-Rear: braking (CCRb)

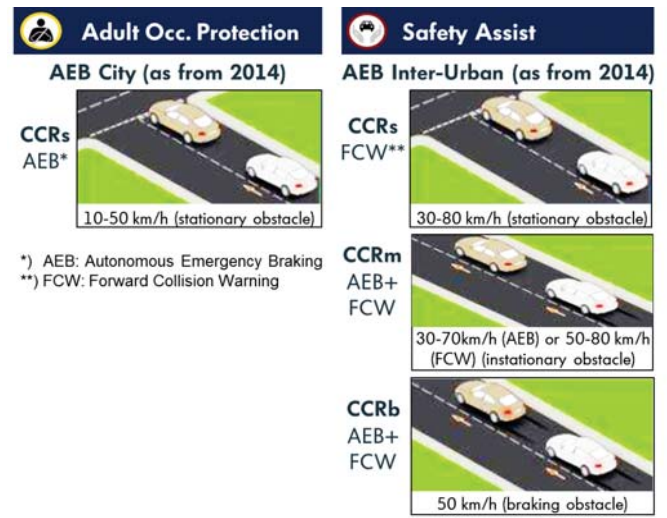

**Figure 1: EuroNCAP's test scenarios separated into different safety columns (based on [8]).**

These scenarios address typical types of crashes in the city and interurban areas. The different test cases will be performed with increasing levels of speed, ranging from 10 to 80 km/h. Test conditions regarding test ground and surface, test and measurement equipment are explicitly specified. The tolerance ranges of the main parameters like velocity, vehicle weight, path deviation and others are described in detail as well.

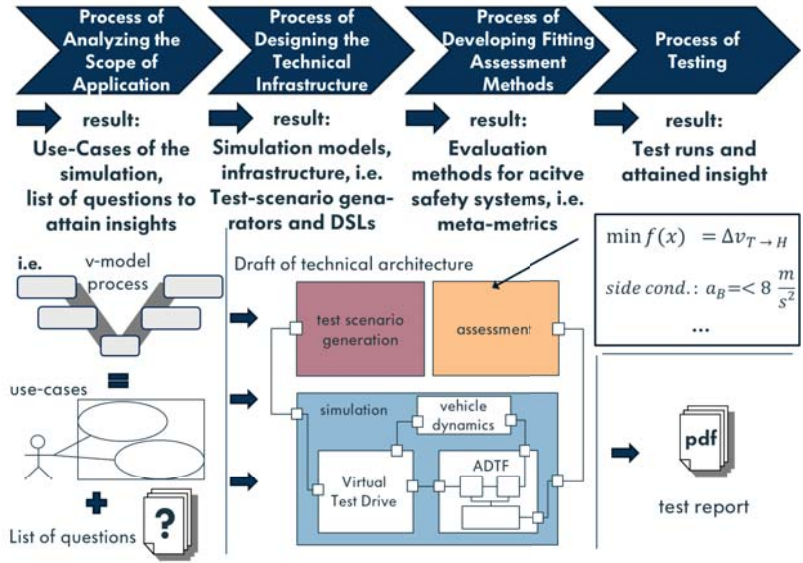

Figure 2: Components of the process to develope an appropriate simulation environment.

## 3.2 Method of Designing a Simulation Environment Efficiently

How the challenges on a series development may be supported effectively by simulation and especially how an assessment may evolve within in the tolerance ranges at EuroNCAP, are discussed in the following.

In order to establish a suitable model of an underlying system that significantly contributes to safety the followed questions must be fundamental to the engineer and answered first: "Which engineering problem should be addressed by the simulation?" and "Which result is expected?". The insight to be attained at the end of the simulation process is crucial for modeling the system and its context because it defines which critical parameters are necessary for an accurate abstraction of reality. An attempt to simulate an exact reality environment carries the risk of process failure due to the amount of details, which only have a marginal influence on the simulation result or the attained insight. To avoid this risk, a method definition is currently in progress that helps an engineer to concentrate on relevant aspects of designing a simulation environment more systematically. The method may be broken down into four different components:

- Process of analyzing the scope of application
- Process of designing the technical infrastructure
- Process of developing fitting assessment methods
- Process of testing

The explanation of the individual components from figure 2 will be now given in the context of analyzing the tolerance range for EuroNCAP's consumer tests.

### 3.2.1 Process of Analyzing the Scope of Application

Two key questions are fundamental to analyze the scope of application of the simulation environment:

1. *Which step of the development process and which tasks should be supported by the simulation?*

2. *What insight should be gained by the simulation?*

The first question aims to gather relevant project circumstances and to clarify the benefit for the engineer. The analysis also illustrates what objectives should be achieved by the simulation in the end, due to the fact that the engineer is sometimes not fully aware of how he could be supported by a simulation environment in the development process.

The second question focuses on the driver assistance systems' field of action. It is important to identify which metrics of the function or which type of coherences between function and field of action should be revealed. That means a specification of the superior objective derived from the task analysis before into the system's processed variables and parameters and their physical background.

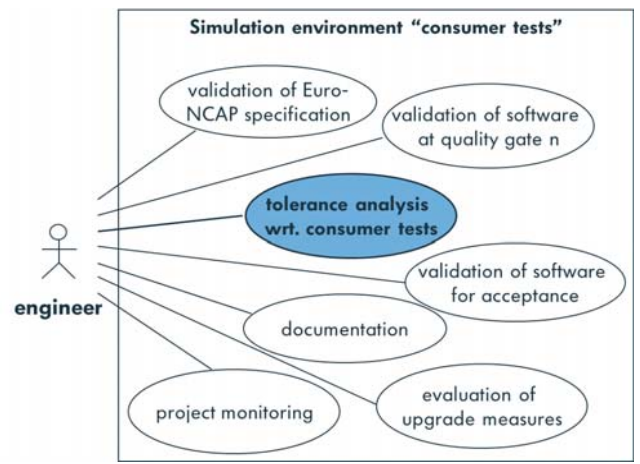

Figure 3: Use case diagram of the project "consumer tests".

It is considered that the use-case diagram as part of the Unified Modeling Language (UML) is a suitable tool to model and to structure these tasks, because the regarding stake-holder can be assigned to the underlying tasks during the development phase [14, 15, 16]. They can often be deduced from internal project documents with a high level of detail. For the reason of non-disclosure the engineer's task has been generalized in this particular case.

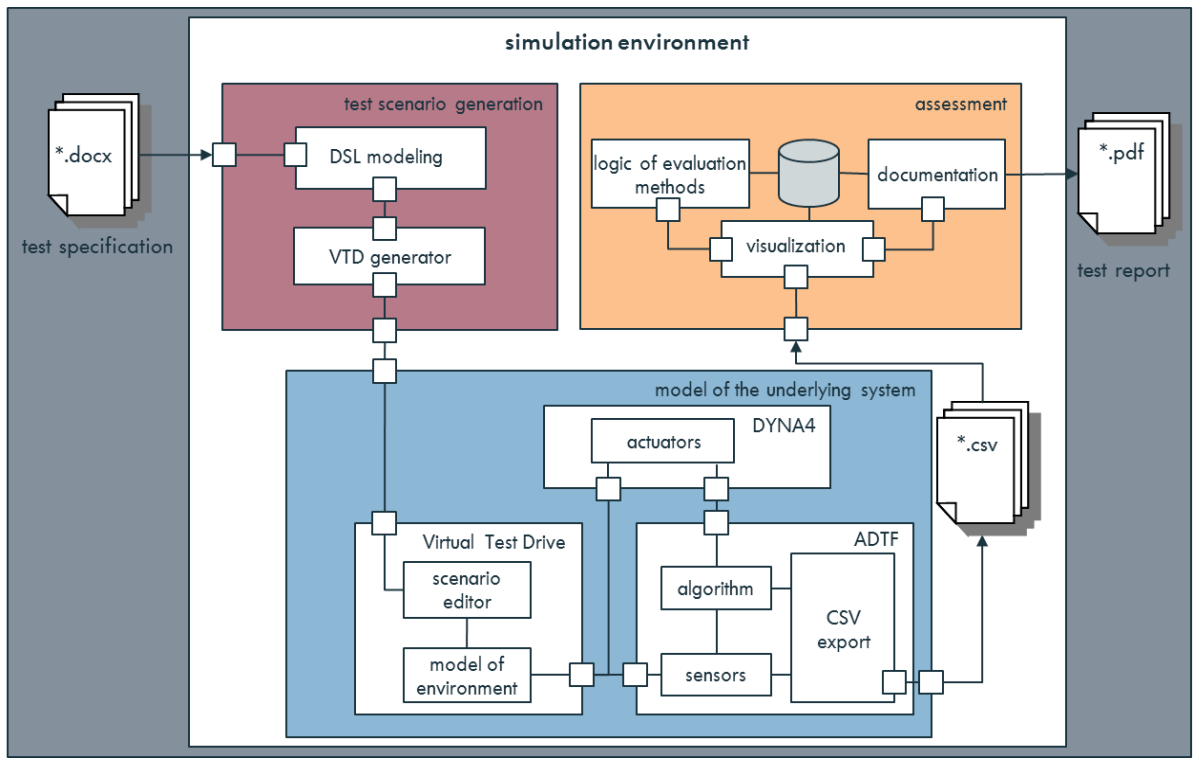

Figure 4: Architecture of the simulation environment.

Regarding the attained insight through simulation, the relevant metrics that should be investigated in view of the tested driver assistance system, must be identified. According to AEB test protocol the remaining speed is the prime metric for assessing the performance of the system. Based on that, other metrics that may be influencial to the remaining speed at impact need to be derived. For eg.: **(i)** longitudinal and lateral positions of the VUT and the target, **(ii)** metrics regarding the braking system itself, **(iii)** the detection performance of the sensor or **(iv)** the conditions of the road surface.

In this regard, not only the system itself but also the static and dynamic context of the system must be intensively analyzed due to the possible variations of other metrics at the same time and thus the potential effect on each other. In the end, the distribution of the possible scoring at EuroNCAP with different test conditions and parameter settings are one of the insights that are expected.

Another insight is expected by identifying those parameters, which have a major influence on the scoring distribution than others. Thus, the engineer would be able to specify requirements more in detail which score at which test case the driver assistance system has to reach and what actions has to be taken to improve the system's performance in that particular test case. He would also be able to evaluate the active safety system at certain quality gates for monitoring aspects.

### 3.2.2 Process of Designing a Technical Infrastructure

The models and the respective software components provide the basis of the technical infrastructure that simulates the underlying system and the series of tests. Beside the modelled system, which generates the synthetical data for sensors, algorithm and actuators, there has to be software components for the generation of test scenarios as well as for the evaluation of the system.

The main attention should be on choosing the right level of abstraction during modelling because the implementation effort can be minimized by concentrating only on the relevant parameters rather than designing a smart representation of reality. The attempt of modeling reality is not only an ambitious challenge, but possibly also leads to an endless loop of specification and realization.

The figure 4 depicts the current design of the underlying architecture, which models a FCW/AEB system. The software components "test scenario generation" and "evaluation" are going to be realized by another project. The test scenarios will be described as a domain-specific language (DSL) developed within the MontiCore framework [7]. In turn, a code generator will create the necessary input data for "Virtual Test Drive" (VTD) [11].

The model of the driver assistance system is realized within the "Automotive Data- and Time-triggered Framework" (ADTF), which communicates with VTD via its Runtime-Data-Bus (RDB) and a special interface within ADTF. The RDB delivers a large number of simulation data [12]. The sensor model, which transforms the relevant object data into a readable format for the FCW/AEB algorithm, simulates in its current implementation an ideal model without signal noise around the object detection. This is based on the assumption that sensor-specific behavior should be dissociated from the behavior of the vehicle dynamics or the algorithm to concentrate on these metrics first.

In a second step, the sensor model will be extended by applying signal noise and other error types, because driver assistance systems depend very much on the sensor's performance capturing the environment. The algorithm itself analyzes the surrounding objects in regard to their criticality for the VUT. That implemented algorithm has a complex structure close to series maturity and is currently used as reference to evaluate other algorithms developed by suppliers for different vehicle projects.

The result will be displayed as a signal that is defined as the grade of a four-step warning-level. This will be the basis for the braking strategy, which passes on the requested value of deceleration to the modeled braking system as part of vehicle dynamic module "DYNA4" [5]. This component is also connected to VTD via RDB and the deceleration request will currently be handled by the ghost driver. In the next stage of this module, there will be a direct connection between ADTF and DYNA4, thus being able to send signals straight to the respective component.

For the purpose of further analysis and evaluation, comma-separated-value files (CSV) are recorded at certain points of the ADTF configuration. Hence, a mechanism is established to use the calculated signal values of the underlying model in other environments for other types of evaluation procedures. A suitable representation of these signal curves is mandatory for a thorough assessment of the simulation runs and should be addressed to the assessment tool as a prime requirement. However, this type of visualization differs substantially from a 3D-graphical representation of the scene. The latter is not necessary for an effective simulation of a driver assistance system. The crucial point of interest is how the different signals can be aggregated to a few metrics or even single one, being able to decide whether a simulation run was successful or not.

### 3.2.3 Process of Developing fitting Assessment Methods

The remaining impact speed between VUT and the target vehicle is the prime metric for the assessment at EuroNCAP. Thus, it has to be verified to what extent variations within the test parameters' tolerance ranges will effect that metric. The distribution of scorable points with different parameter settings and test conditions is significant for the series development to enable the engineer to estimate the the worst case scoring scenario. Furthermore the remaining impact speed can be seen as an optimizable parameter which, with the correct choice of system's parameter settings, warrants the best deceleration.

In order to monitor the development of the software stages realizing the safety function, the method of meta-metrics will be applied and developed further [2]. It is also planned to establish automatic regression tests by automatically detecting and evaluating successful or failed simulation runs. As a result, the engineer will be able to assess the status of a project more precisely.

The objective for the series development is to ensure that the VUT will fulfill the requirements for EuroNCAP's 5-stars rating. By simulating such systems in action, analyzing the tolerance ranges and performing automated regression tests, the achievement of that objective may be supported.

### 3.2.4 Testing Process

The gaining insights will highly depend on the comparability of simulation results and real test data. Due to the modelling of the driver assistance system and its context, abstractions have been made that does not allow a direct comparison of real tests and simulation runs. To ensure comparability, relations between the different results have to be established. How this might be done properly, will be investigated later on during a case study to identify the gap between simulation and reality. The conclusion will be limited to the here outlined use-case and does not allow any generalizations on other use-cases.

Due to the fact that realizing the simulation environment including the infrastructure, the evaluation methods and the test scenario generation is a work in progress, first test results are expected for the first quarter of 2014. An important aspect wrt. equivalence class partitioning (ECP) as a black-box testing method that is discussed more in detail, as ECP might challenge that the tolerance analysis is a necessary task for the series development.

By applying ECP, a representative parameter is chosen to represent a number of input and output data pairs that result in a similar behavior of a software component. Transferring this to the challenge regarding consumer tests, it would mean that the tolerance ranges of the test parameters should not have any significant influence on the algorithm and therefore on the performance of the active safety system. The following example demonstrates that ECP is not sufficient enough for performance tests and a tolerance analysis is still needed.

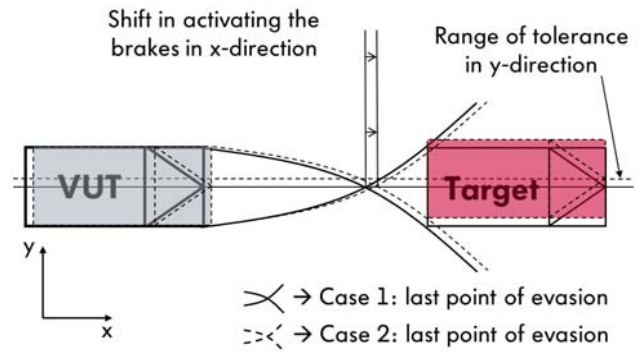

Figure 5: Changes in y-direction may effect the assessment of the situation and thus may influence the impact speed.

In figure 5, it is shown that a shift of the target in y-direction does have an effect on the algorithm triggering the AEB system. If the following equivalence classes are chosen such as [A] behavior without any evasion trajectory, [B] behavior with evasion trajectory "left-handed" and [C] behavior with evasion trajectory "right-handed", selecting one or several representative parameters will not be sufficient for class [A]. The driver assistance system admittedly behaves in a similar way with different input data by braking on target with different positions in y-direction but the ECP does not consider the timing of triggering the braking system, which influences the impact speed and therefore a success-ful consumer test. As a consequence, the tolerance analysis is necessary to estimate particularly how the parameters in combination affect the trigger timing of the algorithm.

## 4. CONCLUSION AND FUTURE WORK

Due to the upcoming assessment of driver assistance systems for EuroNCAP's five-star rating, the systems have to perform on a high quality level. The test protocol allows deviation of testing and system parameters in certain ranges. In fact, these tolerance ranges will influence the impact speed and performance during the tests. Thus, a tolerance analysis must be performed to estimate the worst case scoring scenario of the driver assistance system. It is shown that the method of ECP is not sufficient enough to establish confidence in the behavior of the system from a series development's point of view. In order to face that challenge, a simulation approach as a work in progress is presented.

Future work will focus on further assessment of the simulation runs within the tolerance ranges, establishment of a sensor model with a variable signal noise to identify the influence of sensor-specific behavior. A case study including a comparison of simulation runs with real test runs is also planned as future work.